\documentclass[aps,prd,eqsecnum,12pt,showpacs,preprintnumbers,nofootinbib]{revtex4}

\def\nbox#1#2{\vcenter{\hrule \hbox{\vrule height#2in
\kern#1in \vrule} \hrule}}
\def\sq{\,\raise.5pt\hbox{$\nbox{.09}{.09}$}\,}
\def\sqb{\,\raise.5pt\hbox{$\overline{\nbox{.09}{.09}}$}\,}

\newcommand{\bea}{\begin{eqnarray}}
\newcommand{\eea}{\end{eqnarray}}
\newcommand{\be}{\begin{equation}}
\newcommand{\ee}{\end{equation}}
\newcommand{\bes}{\begin{subequations}}
\newcommand{\ees}{\end{subequations}}

\newcommand{\eq}{\begin{eqnarray}}
\newcommand{\qed}{\end{eqnarray}}

\newcommand{\G}{\ensuremath{{G_0}^0}}
\newcommand{\T}{\ensuremath{\left< {T_0}^0 \right>}}
\newcommand{\ha}{\ensuremath{{{}^{\mbox{\tiny (1)}\!}{H_0}^0}}}
\newcommand{\hb}{\ensuremath{{{}^{\mbox{\tiny (3)}\!}{H_0}^0}}}
\newcommand{\R}{\ensuremath{\left< \rho \right>}}

\usepackage{amssymb,amsmath}
\usepackage{graphicx}
\usepackage{graphics}

\begin{document}
\
\title{Effects of Quantized Scalar Fields in Cosmological Spacetimes with Big Rip Singularities}
\author{Jason D. Bates}
\email{batej6@wfu.edu}

\author{Paul R. Anderson}
\email{anderson@wfu.edu}

\affiliation{Department of Physics,\\
Wake Forest University, \\
Winston-Salem, NC 27109 USA}
\begin{abstract}
\vskip .3cm
Effects of quantized free scalar fields
in cosmological spacetimes with Big Rip singularities are investigated.  The energy densities for
these fields are computed at late times when the expansion is very rapid.  For the massless
minimally coupled field it is shown that an attractor state exists in the sense that, for a large class of states, the energy
density of the field asymptotically approaches the energy density it would have if it was in the attractor state.
Results of numerical computations of the energy density for the massless minimally coupled field
and for massive fields with minimal and conformal couplings to the scalar curvature are presented.
For the massive fields the energy density is seen to always asymptotically approach that of the
corresponding massless field.  The question of whether the energy densities of quantized fields
can be large enough for backreaction effects to remove the Big Rip singularity is addressed.

\end{abstract}


\maketitle

\vfill\eject

\section{Introduction}
\label{introduction}

Surveys of type Ia supernovae and detailed mappings of the cosmic microwave background provide strong evidence that the universe is accelerating \cite{Supernova}.  To explain this acceleration within the framework of Einstein's theory of General Relativity requires the existence of some form of ``dark energy'' which has positive energy density and negative pressure \cite{DE-Review}.

One common model for dark energy \cite{DE-Review} is to treat it as a pervasive, homogenous perfect fluid with equation of state
$p = w \rho$.  Cosmic acceleration demands that $w < -1/3$, and observations from the Wilkinson Microwave Anisotropy Probe in conjunction with supernova surveys and baryon acoustic
  oscillations measurements place the current value at $w=-0.999^{+0.057}_{-0.056}$~\cite{WMAP}.  Although this is consistent with the effective equation of state for a cosmological constant, $w=-1$, we cannot rule out the possibility that our Universe contains ``phantom energy'', for which $w < -1$.

If $w$ is a constant and $w < -1$ then general relativity predicts that as the universe expands the phantom energy density increases with the result that in a finite amount of proper time
the phantom energy density will become infinite and the universe will expand by an infinite amount.  All bound objects, from clusters of galaxies to atomic nuclei, will become unbound as the Universe approaches this future singularity, aptly called the ``Big Rip''~\cite{Caldwell}.

A simple model of a spacetime with a Big Rip singularity can be obtained by considering a spatially flat Robertson-Walker spacetime, with metric
\eq ds^2 &=& -dt^2 + a^2(t)(dx^2 + dy^2 + dz^2). \label{FRW} \qed
At late times the phantom energy density
 is much larger than the energy density of classical matter and radiation and the solution to the classical Einstein equations is
\eq a(t) \approx a_1 (t_r - t)^{-\sigma},  \label{abigrip} \qed
with
\eq \sigma = -2/\left(3+3w\right) > 0 \;. \qed
Here $a_1$ is a constant and $t_r$ is the time that the Big Rip singularity occurs.  The phantom energy density is
\eq \rho_{ph} &=& \frac{3}{8\pi}\sigma^2 (t_r-t)^{-2}.  \label{rhoph} \qed
Note that both the scale factor and the phantom energy density become infinite at $t=t_r$.

In addition to the singularity in the model described above, other classes of models containing somewhat milder phantomlike future singularities have been identified. Barrow~\cite{Barrow} constructed a class of models called ``sudden singularities'' in which $w$ is allowed to vary as $w \propto (t_r - t)^{\alpha-1}$ for $0 < \alpha < 1$. In these models, the pressure and scalar curvature diverge at time $t_r$ but the energy density and scale factor remain finite. Other types of singular behavior were found in Refs.~\cite{Barrow2, NO2, stefancic}. Nojiri, Odintsov and Tsujikawa~\cite{NOT} came up with a general classification scheme for future singularities.  The strongest singularities are classified as type I and Big Rip singularities fall into this class.  The sudden singularities are examples of type II singularities.  Two other classes, type III and type IV were also identified.  For type III singularities both the energy density $\rho$ and the pressure $p$ diverge at the time $t_r$ but the scale factor $a$ remains finite.
   For type IV singularities, the scale factor remains finite, the energy density $\rho$ and the pressure $p$ go to zero at the time $t_r$, but divergences in higher derivatives of $H = \dot{a}/a$ occur.  Other classification schemes
 for cosmological singularities have been given in Refs.~\cite{cattoen-visser, dabrowski, fernandez-lazkoz}.

At times close to $t_r$ in each of the above cases it is possible that quantum effects could become large and that the backreaction
of such effects could moderate or remove the final singularity.
One way to investigate whether this would occur is to compute the energy density for the quantum fields in the background geometry of a spacetime with a final singularity.
Then a comparison can be made between the phantom energy density and the energy density of the quantized fields.
A second way is to solve the semiclassical backreaction equations to directly see what effects the quantum fields have.

Nojiri and Odintsov~\cite{NO1,NO2} studied the backreaction of conformally invariant scalar fields in the cases of sudden singularities and Big Rip singularities and found that quantum effects could delay, weaken or possibly remove the singularity at late times.  In Ref.~\cite{NOT}  Nojiri, Odintsov and Tsujikawa used a model for the dark energy with an adjustable equation of state to find examples of spacetimes with type I, II, and III singularities.  They then solved the semiclassical backreaction equations and found that the singularities were usually either moderated or removed by quantum effects.

Calder\'{o}n and Hiscock~\cite{Hiscock-Calderon} investigated the effects of conformally invariant scalar, spinor, and vector fields on Big Rip singularities by computing the stress-energy of the quantized fields in spacetimes with constant values of $w$.  Their results depend on the value of $w$ and on the values of the renormalization parameters for the fields.  For values of $w$ that
are realistic for our universe they found that quantum effects serve to strengthen the singularity.
Calder\'{o}n~\cite{Calderon} made a similar computation in spacetimes with sudden singularities and found that whether the singularity is strengthened or weakened depends on the sign of one of the renormalization parameters.

Barrow, Batista, Fabris and Houndjo~\cite{Barrow-Batista} considered models with sudden singularities when a massless,
minimally coupled scalar field is present.  They found that in the limit $t \rightarrow t_r$,
the energy density of this field remains small in comparison with the phantom energy density.  Thus quantum effects are never important in this case.
Batista, Fabris and Houndjo~\cite{Batista} investigated the effects of particle production when a massless minimally coupled scalar field is present
in spacetimes where $w$ is a constant.  To do so they used a state for which  Bunch and Davies~\cite{bunch-davies} had previously computed the
stress-energy tensor.  They found that the energy density of the created particles never dominates over the phantom energy density.

Pavlov~\cite{Pavlov} computed both the number density of created particles and the stress-energy tensor for a conformally coupled massive scalar field for the case in which $w = -5/3$.  It was found that backreaction effects
are not important for masses much smaller than the Planck mass and times which are early enough that the time until the Big Rip occurs is greater than the Planck time.

In this paper we compute the energy densities of both massless and massive scalar fields with conformal and minimal couplings to the scalar curvature in spacetimes with Big Rip singularities
in which the parameter $w$ is a constant.
While our calculations are for scalar fields, it is worth noting that both massive and massless conformally coupled scalar fields can be used to model spin $1/2$ and spin $1$ fields, and in homogeneous and isotropic spacetimes the massless minimally coupled scalar field can serve as a model for gravitons~\cite{Grischuck,Ford-Parker}.

For conformally invariant fields the natural choice of vacuum state in homogeneous and isotropic spacetimes is the conformal vacuum~\cite{b-d-book}.  For all other fields there is usually no natural choice.  However,
it is possible to define a class of states called adiabatic vacuum states which, when the universe is expanding slowly, can serve as reasonable vacuum states~\cite{b-d-book}.  They
can be obtained using a WKB approximation for the mode functions, and they are specified by the order of the WKB approximation. It has
been shown that the renormalized stress-energy tensor for a quantum field is always finite if a fourth order or higher adiabatic vacuum state is chosen.\footnote{In this paper we generalize the definition of an n'th order adiabatic vacuum state to include all states whose high momentum modes are specified by an n'th order WKB approximation but whose other modes can
be specified in any way.}

Here we compute the renormalized
energy densities of the quantum fields in fourth order or higher adiabatic states and investigate their behavior as the universe expands.  One focus is on the differences that occur for the same
field in different states.  We find in all cases considered that the asymptotic behavior of the energy density is always the same for fields with the same coupling to the scalar curvature, regardless of whether they are massless or massive and regardless of what states the fields are in.  Fields with minimal coupling to the scalar curvature have a different asymptotic behavior
than those with conformal coupling.

We also address the question of whether and under what conditions the energy density of the quantized fields becomes comparable to the phantom energy density.  We find that for fields in realistic states for which the energy density of the quantized fields is small compared to that of the phantom energy density at early times, and for spacetimes with realistic values of $w$, there is no evidence that quantum effects become large
enough to significantly affect the expansion of the spacetime until the spacetime curvature is of the order of the Planck scale or larger, at which point the semiclassical approximation
breaks down.

 In Sec. II the quantization of a scalar field in a spatially flat Robertson-Walker spacetime is reviewed along with a method of constructing adiabatic states.
 In Sec. III, the energy density for massless scalar fields with conformal and minimal coupling to the scalar curvature is discussed and a comparison is made with
 the phantom energy density. For the massless minimally coupled scalar field a proof is given that one particular
 state serves as an attractor state in the sense that for a large class of states, the energy density of the field asymptotically approaches the energy density it would have if it was in the attractor state.
  In Sec. IV numerical calculations of the energy density for massive scalar fields with conformal and minimal coupling to the scalar curvature are discussed.
  A comparison is made with both the phantom energy density and the energy density of the corresponding massless scalar field.
  Our main results are summarized and discussed in Sec. V.
  Throughout units are used such that $\hbar = c =  G = 1$ and our sign conventions are those of Misner, Thorne, and Wheeler~\cite{MTW}.

\section{Quantization in Spatially Flat Robertson Walker Spacetimes}
\label{QFTRW}

We consider a scalar field $\phi$ obeying the wave equation
\eq (\Box - m^2 - \xi R)\phi = 0, \qed
where m is the mass of the field and $\xi$ is its coupling to the scalar curvature $R$.  It is convenient for our calculations to use the conformal time variable
\eq \eta = \int^t_{t_r}\frac{d\bar{t}}{a(\bar{t})}. \qed
We expand the field in terms of modes in the usual way \cite{b-d-book}
\eq \phi(\mathbf{x},\eta) &=& \frac{1}{a(\eta)} \int \frac{d^3 k}{(2\pi)^\frac{3}{2}} \left[ a_{\mathbf{k}} e^{i\mathbf{k}\cdot\mathbf{x}} \psi_k(\eta) + a_{\mathbf{k}}^{\dagger} e^{-i\mathbf{k}\cdot\mathbf{x}} \psi_k^*(\eta) \right], \qed
with the creation and annihilation operators satisfying the commutation relations
\begin{eqnarray} [a_{\mathbf{k}},a_{\mathbf{k}'}] &=& [a_{\mathbf{k}}^\dagger,a_{\mathbf{k}'}^\dagger] = 0  \nonumber  \\
  \,  [ a_{\mathbf{k}},a_{\mathbf{k}'}^\dagger ] &=&  \delta(\mathbf{k} - \mathbf{k}') \;.
    \end{eqnarray}
The time dependent part of the mode function satisfies the equation
\eq \psi_k'' + \left[k^2+m^2a^2+6\left(\xi - \frac{1}{6}\right)\frac{a''}{a}\right]\psi_k &=& 0 \;, \label{modeeq} \qed
with a constraint given by the Wronskian condition
\eq \psi_k {\psi_k^*}' - \psi_k^* {\psi_k}' = i. \label{wronskian} \qed
Throughout primes denote derivatives with respect to the conformal time $\eta$.

We restrict our attention to states for which the stress-energy tensor is homogeneous and isotropic.  Thus the stress-energy tensor is uniquely specified by
      the energy density $\R_r = -\T$ and the trace $\left<T\right>$.  The unrenormalized energy density is \cite{bunch}
\eq
\R_u &=&  \frac{1}{4\pi^2 a^4}\int_0^{\infty} dk\,k^2 \left\{  |\psi_k'|^2 \,+ \, \left[k^2 + m^2a^2 - 6\left(\xi - \frac{1}{6}\right)\frac{a'^{\,2}}{a^2}\right]|\psi_k|^2 \right. \nonumber \\
     & & \left. + \, 6\left(\xi - \frac{1}{6}\right)\frac{a'}{a}(\psi_k {\psi_k^*}' + \psi_k^* {\psi_k}') \right\}.
\label{Tu} \qed
Renormalization is to be accomplished through the use of adiabatic regularization~\cite{Parker,P-F,F-P,F-P-H}.  Following the prescription given in Ref.~\cite{Anderson-Eaker} the renormalized energy density is
\begin{subequations}
\eq \R_r = \R_u - \R_d + \R_{an} \label{Tr} \qed
with\footnote{ Note that Eq.(9a) of Ref.~\cite{Anderson-Eaker} has a misprint.  The term on the third
line which is proportional to $m^2$ should be multiplied by a factor of $(\xi-1/6)$.}
\eq
\R_d &=& \frac{1}{4\pi^2 a^4}\int_0^\infty dk\,k^2 \left\{k + \frac{1}{k}\left[\frac{m^2 a^2}{2} - \left(\xi - \frac{1}{6} \right)\frac{3 \, a'^{\,2}}{a^2} \right]  \right\} \nonumber \\
& & + \frac{1}{4\pi^2 a^4}\int_\lambda^\infty dk\,k^2 \left\{\frac{1}{k^3} \left[-\frac{m^4 a^4}{8} - \left(\xi - \frac{1}{6} \right)\frac{3m^2 a'^{\,2}}{2} \right. \right. \nonumber \\
& & + \left. \left. \left(\xi - \frac{1}{6}\right)^2 \ha \frac{a^4}{4} \right] \right\}, \\
\R_{an} &=& \frac{1}{2880\pi^2} \left(-\frac{1}{6}\ha + \hb \right) + \frac{m^2}{288\pi^2}\G \nonumber \\
& & -\frac{m^4}{64\pi^2} \left[ \frac{1}{2} + \log{\left(\frac{\mu^2 a^2}{4\lambda^2}\right)} \right] + \left(\xi - \frac{1}{6} \right) \left\{ \frac{\ha}{288\pi^2} \right. \nonumber \\
& & \left. + \frac{m^2}{16\pi^2}\G \left[3 + \log{\left(\frac{\mu^2 a^2}{4\lambda^2}\right)} \right] \right\} \nonumber \\
& & + \left(\xi - \frac{1}{6} \right)^2 \left\{ \frac{\ha}{32\pi^2} \left[2 + \log{\left(\frac{\mu^2 a^2}{4\lambda^2}\right)} \right] -\frac{9}{4\pi^2} \frac{a'^{\,2} a''}{a^7} \right\}\;,
\label{renorm} \qed
and
\eq
\ha &=& -\frac{36a'''a'}{a^6} + \frac{72a''a'^{\,2}}{a^7} + \frac{18a''^{\,2}}{a^6} \\
\hb &=& \frac{3a'^{\,4}}{a^8} \\
\G &=& -\frac{3a'^{\,2}}{a^4}.
\qed
\end{subequations}

\noindent Note that the value of $\R_r$ is independent of the arbitrary cutoff $\lambda$.
For a massive field $\mu = m$ whereas for a massless field $\mu$ is an arbitrary constant.  The renormalization procedure for the trace $\left<T\right>$ is given in Ref.~\cite{Anderson-Eaker} and follows similar lines.  For the purposes of this paper, we are concerned only with the energy density.

\subsection{WKB Approximation and Adiabatic States}
\label{wkbsec}

A formal solution to the time dependent part of the mode equation can be given using a WKB expansion.  First make the change of variables
\eq \psi_k(\eta) = \frac{1}{\sqrt{2W(\eta)}}\exp{\left[-i\int^{\eta}_{\eta_0} W(\bar{\eta})\,d\bar{\eta} \right]} \label{WKB-mode} \qed
where $\eta_0$ is some arbitrary constant.  Substituting into Eq.~\eqref{modeeq} gives
\eq W = \left[k^2 + m^2a^2 + 6\left(\xi - \frac{1}{6} \right)\frac{a''}{a} - \frac{1}{2}\left(\frac{W''}{W} - \frac{3W'^2}{2W^2}\right)\right]^\frac{1}{2} \label{WKB} \;. \qed

Approximate solutions to the above equation may be obtained using an iterative scheme, where each successive iteration is given by
\eq W_k^{\left( n \right)} =
\left(k^2 + m^2 a^2 + 6\left(\xi - \frac{1}{6} \right)\frac{a''}{a}
- \frac{1}{2}\left[\frac{{{W_k}^{\left( n-2 \right)}}''}{W_k^{\left( n-2 \right)}}
- \frac{3}{2}\left(\frac{{W_k^{\left( n-2 \right)}}'}{W_k^{\left( n-2 \right)}}\right)^2 \right]\right)^{\frac{1}{2}}, \label{WKB-n} \qed
with
\eq W_k^{\left( 0 \right)} \equiv \left(k^2 + m^2a^2\right)^\frac{1}{2}. \label{WKB0} \qed
Taken in conjunction with Eq. \eqref{WKB-mode}, this can be used to define an $n$th-order WKB approximation to the mode equation,
\eq \psi_k^{\left( n \right)}(\eta) = \frac{1}{\sqrt{2W_k^{\left( n \right)}(\eta)}}\exp{\left[-i\int^{\eta}_{\eta_0} W_k^{\left( n \right)}(\bar{\eta})\,d\bar{\eta}\right]}. \label{WKB-mode-n} \qed
In general, this approximation is valid when $k$ is large and/or the scale factor varies slowly with respect to $\eta$.  In these cases it will differ from the true solution by terms of higher order than $n$.

Exact solutions of the mode equation can be specified by fixing the values of $\psi_k$ and $\psi_k'$ at some initial time $\eta_0$ in such a way that
the Wronskian condition~\eqref{wronskian} is satisfied.
One way to do this is to use an $n$th-order WKB approximation to generate these values.  For example if~\eqref{WKB-mode-n} is used then
\bea \psi_k (\eta_0) &=& \psi_k^{\left( n \right)} (\eta_0) = \frac{1}{\sqrt{2W_k^{\left( n \right)}(\eta_0)}}   \nonumber \\
  \psi_k' (\eta_0) &=& \psi_k'^{\left( n \right)} (\eta_0) = -i \sqrt{\frac{W_k^{\left( n \right)}(\eta_0)}{2}} \;. \label{adstate} \eea
Note that, while it is acceptable to include derivatives of $W_k^{\left( n \right)}$ in the expression for $\psi_k'$, it is not
necessary because such terms are of order $n+1$.

If Eq.~\eqref{adstate} is substituted into Eq.~\eqref{Tu} and then the quantity $\T_r$ is computed, it is found that there
are state dependent ultraviolet divergences in general for zeroth-order and second-order adiabatic states.  There are no such
state dependent ultraviolet divergences for fourth- or higher-order adiabatic states.  This is true for all times because
in a Robertson-Walker spacetime an $n$th-order adiabatic state always remains an $n$th-order adiabatic state~\cite{b-d-book}.

In what follows we are interested in investigating a large class of states.  Thus as mentioned in footnote 1 in the Introduction
we generalize the definition of an $n$th-order adiabatic vacuum state to include all states whose high momentum modes are specified in the above way
by an $n$th-order WKB approximation but whose other modes can
be specified in any way.  However, we shall not consider states in this paper which result in an infrared divergence of the energy density.

\section{Massless Scalar Fields}
\subsection{The Massless Conformally-Coupled Scalar Field}

Setting $\xi=1/6$ and $m=0$ in Eq.~\eqref{modeeq} gives the mode equation for the conformally coupled scalar field
\eq \psi_k'' + k^2\psi_k &=& 0\; \qed
which has solutions
\eq \psi_k &=& \alpha \frac{e^{-ik\eta}}{\sqrt{2k}} + \beta \frac{e^{ik\eta}}{\sqrt{2k}}\;. \qed
The conformal vacuum is specified by the state with $\alpha = 1$ and $\beta = 0$.  Following the renormalization procedure given in Eq.~\eqref{Tr}, we note that the contribution from the modes $\R_u$ is exactly canceled by the counter-terms in $\R_d \;$.  Thus, the renormalized energy density is given by
\eq \R_r = \R_{an} = \frac{1}{2880\pi^2} \left(-\frac{1}{6}\ha + \hb \right). \label{ccm0} \qed

To determine the point at which quantum effects become important one can compare the energy density of the quantum field with the phantom energy density.
This can be done analytically by using Eqs.~\eqref{a} and~\eqref{gamma} to write Eq.~\eqref{ccm0} in terms of the scalar curvature as
\eq \R_r &=& \frac{1}{34560\pi^2}\frac{27w^2 + 18w - 5}{(1-3w)^2}R^2. \label{ccm0-R} \qed
The phantom energy density in terms of the scalar curvature is given by
\eq \rho_{ph} &=& \frac{1}{8\pi(1-3w)}R \;. \label{phantom-R} \qed
Since $R$ diverges at the time $t_r$ it is clear that eventually the energy density of the conformally invariant field becomes comparable to and then
larger than the phantom energy density.  The two densities are equal at the time when
\eq R &=& 4320\pi \frac{1-3w}{27w^2+18w-5}\;.  \label{R0} \qed
For the semiclassical approximation to be valid we need the spacetime curvature to be much less than the Planck scale or $R \ll 1$.
  From Eq.\eqref{R0} one finds the two densities are equal when $R=1$ if
\eq
 w  = -\,\frac{1}{9}\left(3+2160\pi+2\sqrt{6+6480\pi+1166400\pi^2}\right) \approx -1500 \;.
\qed
 Thus for any physically realistic value of $w$ the energy density of the scalar field will remain small compared to the phantom energy density until the scalar curvature is above the Planck scale.  For example, for $w=-1.25$, the energy density of the scalar field equals the phantom energy density at $R \approx 4400$.

\subsection{The Massless Minimally-Coupled Scalar Field}
\label{massless}

Setting $m = \xi = 0$ in \eqref{modeeq} gives the mode equation for the massless minimally coupled scalar field,
\eq \psi_k'' + \left(k^2 - \frac{a''}{a}\right)\psi_k &=& 0. \qed
In terms of the conformal time $\eta$ the scale factor is
\eq a(\eta) = a_0(-\eta)^{-\gamma}. \label{a} \qed
with
\eq \gamma = -\,\frac{2}{1+3w} >0  \label{gamma} \qed
and $a_0$ a positive constant.

The general solution to this equation is given in terms of Hankel functions,
\eq \psi_k(\eta) = \frac{\sqrt{-\pi \eta}}{2}\left(\alpha H^{(1)}_{\gamma +(1/2)}(-k\eta) + \beta H^{(2)}_{\gamma+(1/2)}(-k\eta)\right) \;. \label{Hankel-modes} \qed
Substitution into the Wronskian condition~\eqref{wronskian} gives the constraint
\eq |\alpha|^2 - |\beta|^2 = 1 \;. \label{wronskian2}  \qed

A special case is the solution
 with $\alpha = 1$ and $\beta = 0$.  The stress-energy tensor was computed analytically for the state specified by this solution by Bunch and Davies~\cite{bunch-davies}.  Therefore, in what follows, we will refer to this state as the Bunch-Davies state.  Bunch and Davies found that\footnote{In the process of our calculation, we discovered a misprint in Eq.\ (3.36) of Ref.~\cite{bunch-davies} which
 is repeated in Eq.\ (7.52) of Ref.~\cite{b-d-book}.  The final term in both equations should be multiplied by a factor of 2.}

\eq
\R_r^{BD} &=& \frac{1}{2880\pi^2} \left(-\frac{1}{6}\, \ha + \hb \right) \nonumber \\
& & -\,\frac{1}{1152\pi^2} \ha \left[\log \left( \frac{R}{\mu^2} \right) + \psi(2 + \gamma) + \psi(1 - \gamma) +\,\frac{4}{3} \right] \nonumber \\
& & +\frac{1}{13824\pi^2} \left[ -24\Box R + 24R {R_0}^0 + 3R^2 \right] - \frac{R}{96\pi^2 a^2 \eta^2}  \nonumber \\
&=& -\frac{1}{69120\pi^2}\frac{351w^2+54w-65}{(3w-1)^2}R^2 \nonumber \\
& & +\,\frac{1}{256\pi^2}\frac{(w+1)}{(3w-1)}R^2\left[\log \left( \frac{R}{\mu^2} \right) + \psi\left(\frac{3+3w}{1+3w}\right)  + \psi\left(\frac{6 w}{1+3w}\right)+\frac{4}{3} \right]. \label{BD}
\qed
The mass scale $\mu$ in the above expression is an arbitrary constant and $\psi$ is the digamma function.

As in the case for the massless conformally coupled scalar field, one can compare Eq.~\eqref{BD} with Eq.~\eqref{phantom-R} and see that eventually the energy density of the scalar
field will become comparable to and then greater than the phantom energy density.  A graphical analysis shows that for $w \stackrel{>}{_\sim} -140$  the two energy densities become comparable at a
time when $R \stackrel{>}{_\sim} 1$.
For example, if $w=-1.25$ and $\mu=1$, the energy density of the scalar field equals the phantom energy density at $R \approx 180$.

The Bunch Davies state turns out to be an attractor state in the sense that the energy density for all homogeneous and isotropic fourth order or higher adiabatic states approaches $\R_r^{BD}$. This type of behavior was found in Ref.~\cite{anderson-et-al} in de Sitter space for scalar fields in the Bunch-Davies state and we use the same type of argument here as was used in that paper to establish it. First note that since the renormalization counter terms are the same for any choice of $\alpha$ and $\beta$ in Eq.~\eqref{Hankel-modes}, the renormalized energy density can be written in terms of the energy density of the Bunch-Davies state plus remainder terms.  By substituting Eq.~\eqref{Hankel-modes} into Eq.~\eqref{Tu}, setting $m=\xi=0$, and using the constraint given in Eq.~\eqref{wronskian2}, one finds that for this class of states the energy density can be written as
\eq
\R_r &=& \R_r^{BD} + I(\eta),
\qed
with
\eq
I(\eta) &=& \frac{(-\eta)}{8\pi a^4}\int_0^{\infty} dk\,k^4  \left\{ |\beta|^2 \left[ \left| H_{\gamma-\frac{1}{2}}^{(1)} (-k\eta) \right|^2 + \left| H_{\gamma+\frac{1}{2}}^{(1)} (-k\eta) \right|^2 \right] \right. \nonumber \\
    & &  + \left. Re \left( \alpha\beta^* \left[ \left(H_{\gamma-\frac{1}{2}}^{(1)} (-k\eta) \right)^2 + \left(H_{\gamma+\frac{1}{2}}^{(1)} (-k\eta)\right)^2 \right] \right) \right\}. \label{BDremainder}
\qed

The integral in the above expression can be split into three parts,
\eq I(\eta) = I_1(\eta) + I_2(\eta) + I_3(\eta) \nonumber
\qed
\begin{subequations}
\eq
I_1(\eta) &=&  \frac{(-\eta)}{8\pi a^4}\int_0^\lambda dk\,k^4 \{...\} \label{I3eq-a}  \\
I_2(\eta) &=&  \frac{(-\eta)}{8\pi a^4}\int_\lambda^{-Z/\eta} dk\,k^4 \{...\} \label{I3eq-b}  \\
I_3(\eta) &=&  \frac{(-\eta)}{8\pi a^4}\int_{-Z/\eta}^\infty dk\,k^4 \{...\} \label{I3eq-c}
 \qed
\end{subequations}
where $\lambda$ and  $Z$ are positive constants.  To have a fourth order adiabatic state, $|\beta|$ must fall off faster than $c_1/k^4$ for large enough values of $k$ and any positive constant $c_1$,  thus $\lambda$ is chosen such that $|\beta| < 1/k^4$ when $k \geq \lambda$.  The constant $Z$ is chosen such that $0 < Z << 1$.

To evaluate the first integral, we use the series expansion
\eq H^{(1)}_\nu(x) = - i\, \frac{\Gamma(\nu)}{\pi}\left(\frac{x}{2}\right)^{-\nu} + O(x^{-\nu+2}) + O(x^{\nu}) \; . \label{Hankel-expansion} \qed
This is valid when $x = -k\eta \leq Z$.  We note that at late enough times $Z > -\lambda\eta $ for any choice of $Z$ and $\lambda$.  Using the expansion~\eqref{Hankel-expansion} in Eq.~\eqref{BDremainder}, one finds to leading order that
\eq I_1(\eta) &=& \frac{2^{2\gamma} \left[\Gamma(\gamma+\frac{1}{2})\right]^2}{4\pi^3 a^4_0}(-\eta)^{2\gamma} \int_0^\lambda dk\, \left\{ k^{3-2\gamma}\left[ \left|\beta\right|^2
\right. \right. \nonumber \\ && \left. \left. - Re\left(\alpha\beta^*\right)\right] + O\left[(- k \eta)^{2}\right] + O\left[(- k \eta)^{2 \gamma + 1} \right] \right\} \;. \qed
For states with no infrared divergences the integral is finite, so we find that $I_1(\eta) \rightarrow 0$ as $(-\eta) \rightarrow 0$.

For the second integral, due to the time dependence in the upper limit cutoff, we find that all orders of the expansion given in Eq.~\eqref{Hankel-expansion} contribute, so a different treatment is necessary.  Using the fact that for $k \geq \lambda$ we have $|\beta|^2 < 1/k^8$ and $|\alpha\beta^*| < c_2/k^4$ for some positive constant $c_2$, we can put the following upper bound on the integral in Eq.~\eqref{I3eq-b},
\eq |I_2(\eta)| &<&  \frac{(-\eta)}{8\pi a^4}\int_\lambda^{-Z/\eta} dk\, \left\{ \left( \frac{1}{k^4} + c_2 \right)\left[ \left(J_{\gamma+\frac{1}{2}}(-k\eta)\right)^2 + \left( J_{\gamma-\frac{1}{2}}(-k\eta) \right)^2 \right. \right. \nonumber \\
& & + \left. \left. \left( Y_{\gamma+\frac{1}{2}}(-k\eta)\right)^2 + \left(Y_{\gamma-\frac{1}{2}}(-k\eta)\right)^2 \right] \right. \nonumber \\
& & \left. + 2c_2 \left[ \left|J_{\gamma+\frac{1}{2}}(-k\eta) \, Y_{\gamma+\frac{1}{2}}(-k\eta) \right| + \left| J_{\gamma-\frac{1}{2}}(-k\eta) \, Y_{\gamma-\frac{1}{2}}(-k\eta)\right| \right] \right\} \; . \label{I2}
\qed
Assuming $\gamma \neq \frac{1}{2}$, one can use the standard series expansions for Bessel functions to show that the terms in the integrand are all of the form
\eq \displaystyle (-\eta)^{4 \gamma + 1}  \int_\lambda^{-Z/\eta} dk\, \left( \frac{\tilde{c_1}}{k^4} + \tilde{c_2} \right) \sum_{n,n'=0}^\infty \frac{(-1)^{n+n'}}{n!n'!} \frac{\left(-k\eta\right)^{(\alpha_1 + \alpha_2)p + 2n + 2n'}}{\Gamma(\alpha_1p+n+1)\Gamma(\alpha_2p+n'+1)} \label{g-term}
\qed
where $\alpha_1$ and $\alpha_2$ independently take on the values $\pm 1$, $p$ takes on the values $\gamma\pm\frac{1}{2}$,   and $\displaystyle \tilde{c_1}$ and  $\displaystyle \tilde{c_2}$ are constants.
After integration, Eq.~\eqref{g-term} becomes
\eq \displaystyle & & (-\eta)^{4\gamma} \sum_{n,n'=0}^\infty \frac{(-1)^{n+n'}}{n!n'!} \frac{1}{{\Gamma(\alpha_1p+n+1)\Gamma(\alpha_2p+n'+1)}} \nonumber \\
 & &  \times \left[ \left( \frac{(-\eta)^4 \tilde{c_1}}{Z^4((\alpha_1 + \alpha_2)p + 2n + 2n'-3)} + \frac{\tilde{c_2}}{(\alpha_1 + \alpha_2)p + 2n + 2n'+1} \right) Z^{(\alpha_1 + \alpha_2)p + 2n + 2n'+1}\right. \nonumber \\
 & & \left. - \left( \frac{\tilde{c_1}}{\lambda^4((\alpha_1 + \alpha_2)p + 2n + 2n'-3)} \right. \right.  \nonumber \\
& & \left. \left. + \frac{\tilde{c_2}}{(\alpha_1 + \alpha_2)p + 2n + 2n'+1} \right)
   (-\lambda \eta)^{(\alpha_1 + \alpha_2)p + 2n + 2n'+1} \right] \label{g-term-int}
\qed
Note that the above series converges.  The smallest power of $(-\eta)$ is $(-\eta)^{2\gamma}$ and occurs when $\alpha_1=\alpha_2 = -1$ and $p=\gamma+\frac{1}{2}$.  It comes from the $n = n' = 0$ term.  Thus, $I_2(\eta) \rightarrow 0$ as $(-\eta) \rightarrow 0$.

In the case $\gamma = \frac{1}{2}$, we must use the appropriate series expansions for $Y_{0}(-k\eta)$ and $Y_{1}(-k\eta)$, so Eq.~\eqref{g-term-int} will contain different terms.  However the argument is similar and the result that $I_2(\eta) \rightarrow 0$ as $(-\eta) \rightarrow 0$ is the same.

For the third integral, using $|\beta|^2 < 1/k^8$ and $|\alpha\beta^*| < c_2/k^4$, and changing variables to $x = -k\eta$ gives the bound
\eq |I_3(\eta)| &\leq& \frac{(-\eta)^{4\gamma+4}}{8\pi a_0^4} \int_Z^\infty dx\, \frac{1}{x^4} \left[ \left| H_{\gamma-\frac{1}{2}}^{(1)} (x) \right|^2 + \left| H_{\gamma+\frac{1}{2}}^{(1)} (x) \right|^2 \right] \nonumber \\
& & + \frac{ c_2(-\eta)^{4\gamma}}{8\pi a_0^4} \int_Z^\infty dx\, \left| \left(H_{\gamma-\frac{1}{2}}^{(1)} (x)\right)^2  + \left(H_{\gamma+\frac{1}{2}}^{(1)} (x) \right)^2\right| \;. \label{I3}
\qed
By noting that $\left| H_{\nu}^{(1)} (x) \right| \sim 1/\sqrt{x}$ at large $x$, it is clear that these integrals are finite.  Further, each integral is independent of $(-\eta)$, so the contribution from this piece goes like some constant times $(-\eta)^{4\gamma}$.  Thus, $I(\eta) \rightarrow 0$ as $(-\eta) \rightarrow 0$ and this completes our proof that $\R_r^{BD}$ is an attractor state.

To demonstrate this behavior numerically, we compare the energy density from Eq. \eqref{BD} which occurs for $\alpha = 1$ and $\beta = 0$ to the renormalized energy density of a different fourth order adiabatic state.  One way to specify a fourth order adiabatic state is to use a fourth order WKB approximation as discussed in Sec.~\ref{wkbsec}.  However, for the spacetimes we are considering, using a fourth order WKB approximation in Eqs.~\eqref{WKB-mode-n} and~\eqref{adstate} in the limit $k \rightarrow 0$ results in an infrared divergence of $\R$.  To avoid such a divergence, we use a zeroth order WKB approximation for the modes with $0 \le k \le k_0\;,$ for some positive constant $k_0$.  According to the generalized definition we give in Sec.~I,
these are still fourth order adiabatic states.  A given state is specified by the value of $k_0$ and the time $\eta_0$ at which the matching in Eq.\eqref{adstate} is done.

\begin{figure}
\vskip -0.2in \hskip -0.4in
\includegraphics[width=6in,clip]
{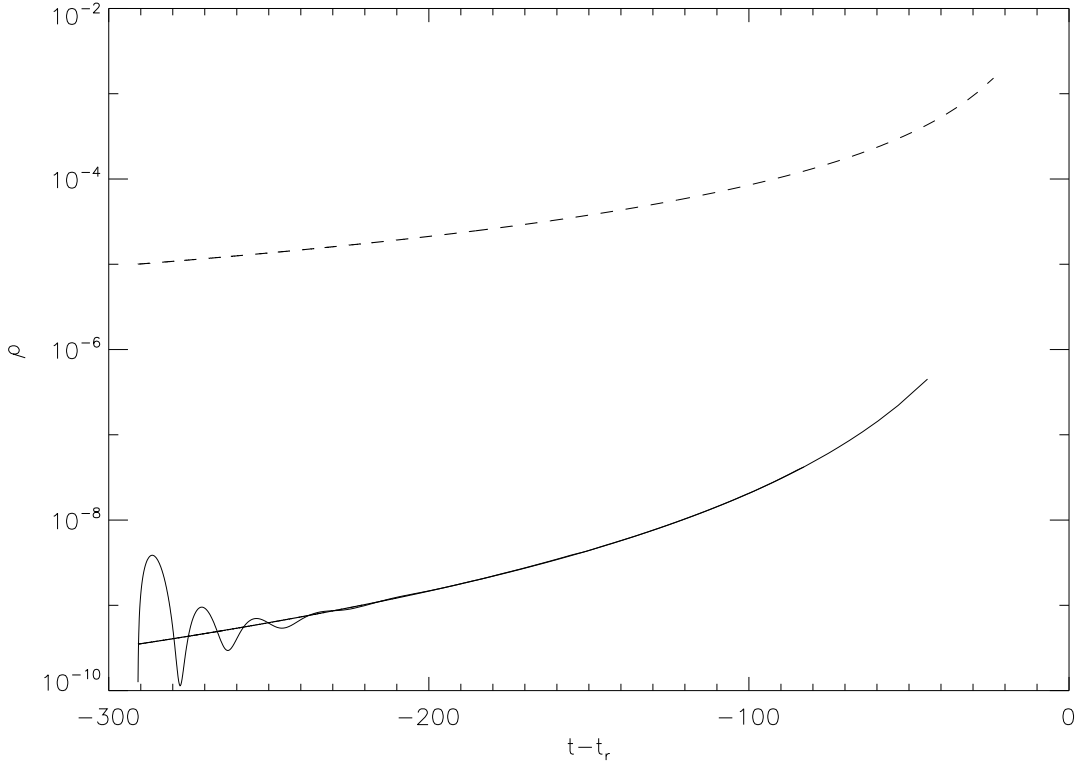}
\vskip -0.2in
\caption{\label{fig:cap1} The dashed line represents the phantom energy density $\rho_{ph}$ in the case that $w=-1.25$. The two solid lines represent the energy density $\R_r$ for two states of the massless minimally coupled scalar field.  The central line is the Bunch Davies attractor state.  The oscillating line is a fourth order adiabatic state with $k_0=0.2$ and $\eta_0 = -100$.  For both states the
mass scale $\mu = 1$ has been used.}
\label{fig1}
\end{figure}

As shown in Fig.~\ref{fig1}, the energy density of the state specified by choosing $k_0=0.2$ and $\eta_0=-100$, approaches that of the Bunch Davies state as expected.
A number of other numerical calculations were done for various values of $k_0$ and $\eta_0$ with the same qualitative result.  Note that
for all the numerical calculations that were done for this paper, $w = -1.25$ and $a_0 \approx 22.6$.

It is of course possible to construct states for which, at times close to $\eta_0$, the energy density is large compared to the phantom energy density.  However, there is no
reason to expect that the energy density of quantized fields at early times should be comparable to that of the phantom energy density since that is not the case today.
Thus our numerical results provide evidence that if the initial energy density is small compared to the phantom energy density at times close to $\eta_0$ then, for
realistic values of $w$, it will remain small during the period when the semiclassical approximation should be valid.

\section{Massive Scalar Fields}
\subsection{Massive Conformally Coupled Scalar Fields}

\begin{figure}
\vskip -0.2in \hskip -0.4in
\includegraphics[width=6in,clip]
{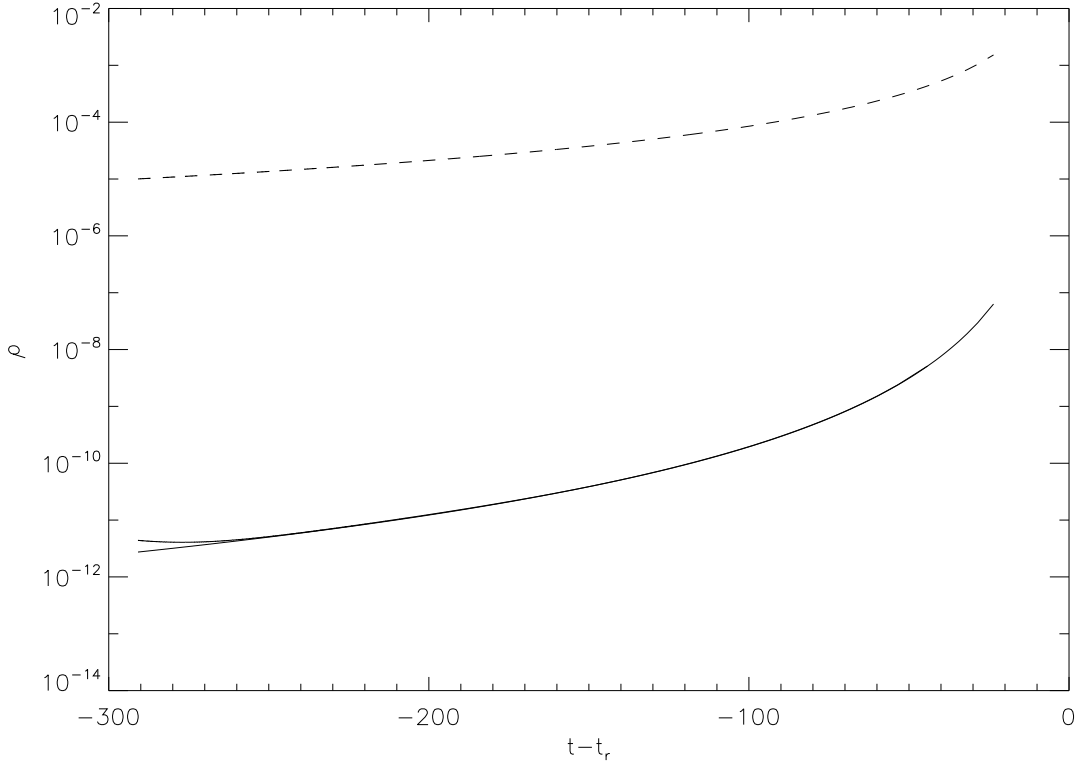}
\vskip -0.2in
\caption{\label{fig:cap2l}The dashed line represents the phantom energy density $\rho_{\rm ph}$ in the case that $w=-1.25$.  The upper solid line is $\R_r$ for a massive ($m=0.005$) conformally coupled scalar field in the fourth order adiabatic state with $k_0=0.1$ and $\eta_0=-100$; the lower one is $\R_r$ for the massless conformally coupled scalar field.}
\label{fig2}
\end{figure}

Setting $\xi=1/6$ in Eq.~\eqref{modeeq} gives the mode equation for a massive conformally coupled scalar field,
\eq \psi_k'' + \left(k^2+m^2a^2\right)\psi_k &=& 0\;. \qed
In this case, we do not have an analytic solution for the mode equation and there is no obvious choice for the vacuum state.  Instead, we choose our initial state using the method outlined near the end of Section~\ref{massless} and compute solutions to the mode equation and the energy density numerically.  What we find is that there is an initial contribution to the energy density from the modes of the field in addition to the contribution from Eq.~\eqref{ccm0}; however, as the Universe expands this contribution redshifts away and the energy density approaches the energy density of the massless conformally coupled scalar field.  An example is shown in Fig.~\ref{fig2} for a field with mass $m = 0.005$ in a state specified by $k_0 = 0.1$ and $\eta_0 = -100$.  We have tested this for several different initial states and different masses and found consistent behavior in all cases.

As with the massless minimally coupled scalar field, it is possible to construct states with a large initial contribution to the energy density and these states can be ruled out
by the same argument as given in Section~\ref{massless}.  Our numerical results provide evidence
that if the energy density of the quantized field is small compared to the phantom energy density at times close to $\eta_0$, then, for
realistic values of $w$, it will remain small during the period when the semiclassical approximation should be valid.

\subsection{Massive Minimally-Coupled Scalar Fields}

The last case we consider is that of massive, minimally coupled scalar fields, which obey the mode equation
\eq \psi_k'' + \left(k^2+m^2a^2 - \frac{a''}{a}\right)\psi_k &=& 0. \qed
Again, we choose our initial state using the method outlined near the end of Section~\ref{massless} and evolve the modes forward in time numerically.

\begin{figure}
\vskip -0.2in \hskip -0.4in
\includegraphics[width=6in,clip]
{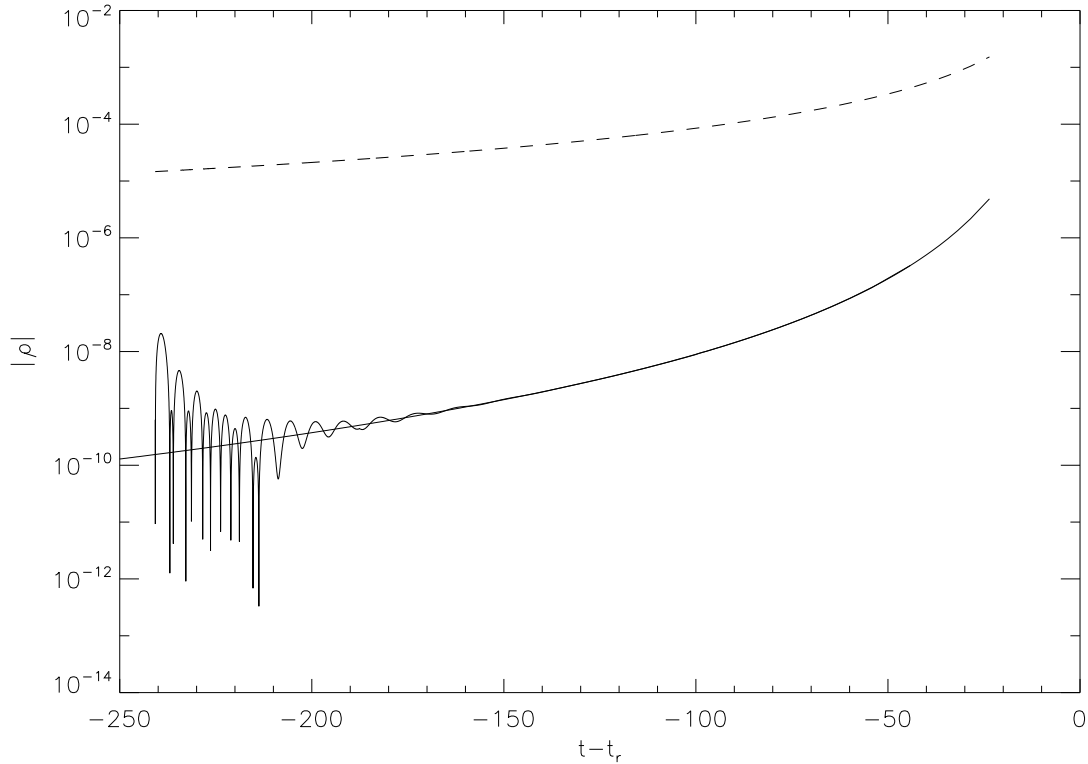}
\vskip -0.2in
\caption{\label{fig:cap3} The dashed line represents the phantom energy density $\rho_{\rm ph}$ in the case that $w=-1.25$.  The central solid line is the energy density $\R_r$ for a massless minimally coupled scalar field in the Bunch Davies state with mass scale $\mu=0.001$.  The oscillating line is the energy density for a massive ($m=0.001$) minimally coupled scalar field in the fourth
 order adiabatic state with $k_0=1$ and $\eta_0=-50$.  Note that the quantity $|\rho|$ rather than $\rho$ has been plotted since the energy density of the massive field is negative at $\eta = \eta_0$, oscillates between positive and negative values for a period of time, and then becomes positive definite at late times.  Note that, except at the initial time, the times where the curve is nearly vertical when it reaches a lower limit are times at which the energy density goes through zero, which is $-\infty$ on the scale of this plot.  Thus the curve should really extend down to $\infty$ at these points.}
\label{fig3}
\end{figure}

Qualitatively, the behavior here is similar to the behavior found for the massless, minimally coupled field in a fourth order adiabatic vacuum state.  At late times, the contribution from the mass terms is small compared to other terms and the energy density approaches that of the massless minimally coupled scalar field in the Bunch-Davies vacuum state.  We note that there is an arbitrary mass scale $\mu$ present in the renormalized energy density~\eqref{BD} for the massless scalar field; the convergence shown in Fig.~\ref{fig3} occurs when $\mu$ is set equal to the mass of the massive field.  At very late times the value of $\mu$ does not affect the leading order behavior of $\R$ for the massless minimally coupled scalar field and thus,
for any value of $\mu$, the energy density of the massive field asymptotically approaches that of the massless field in all cases considered.

\section{Summary and Conclusions}

We have computed the energy densities of both massless and massive quantized scalar fields with conformal and minimal coupling to the scalar curvature in spacetimes with big rip singularities.
We have restricted attention to states which result in a stress-energy tensor which is homogeneous and isotropic and free of ultraviolet and infrared divergences.  For the numerical computations
we have further restricted attention to states for which, near the initial time of the calculation, the energy density of the quantum field is much less than that of the phantom field.

For the massless minimally coupled scalar field we have shown that the energy density for the field in any fourth order or higher adiabatic state
for which the stress-energy tensor is homogeneous, isotropic, and free of infrared divergences, always asymptotically approaches the energy density which this field has in the Bunch-Davies state.  In this sense the Bunch-Davies state is an attractor state.

For massive minimally coupled scalar fields numerical computations have been made of the energy density for different fourth order adiabatic states and in every case
considered the energy density approaches that of the Bunch-Davies state for the massless minimally coupled scalar field at late times.
For conformally coupled massive scalar fields numerical computations have also been made of the energy density for different fourth order adiabatic states.  In each case considered the energy density asymptotically approaches that of the massless conformally coupled scalar field in the conformal vacuum state.  Thus it appears that the
asymptotic behavior of the energy density of a quantized scalar field in a spacetime with a big rip singularity depends only upon the coupling of the field to the scalar curvature
and not upon the mass of the field or which state it is in, at least within the class of states we are considering.

Analytic expressions for the energy densities of both the massless conformally coupled scalar field in the conformal vacuum state and the massless minimally coupled scalar field
in the Bunch-Davies state, in spacetimes with big rip singularities have been previously obtained~\cite{b-d-book, bunch-davies} and are shown in Eqs.~\eqref{ccm0} and \eqref{BD}.
To investigate the question of whether backreaction effects are important in these cases, the energy density of the scalar field can be compared to the phantom energy density to see if there
is any time at which
they are equal or at least comparable.  Then one can determine whether the semiclassical approximation is likely to be valid at this time by evaluating
the scalar curvature and seeing whether or not it is well below the Planck scale.  We have done this and find that for the conformally coupled field the two energy densities
are equal at the point when the scalar curvature is at the Planck scale if $w \sim -1500.$  For the minimally coupled scalar field this combination occurs if $w \sim -140$.
Thus for $w \ll -1500$ for the conformally coupled field and $w \ll -140$ for the minimally coupled field, one expects that backreaction effects may be important at times when the scalar curvature is well below the Planck scale.
However, the values of $w$ which satisfy these constraints are completely ruled out by cosmological observations.

Another way in which the energy density of a quantum field can be comparable to the phantom energy density at scales well below the Planck scale is to construct a state for which this is true.  There is no doubt that such states exist. The analytic and numerical evidence we have is that over long periods of time the energy density of a conformally or minimally coupled scalar field in such a state would
decrease and at late enough times become comparable to that of a massless field in the conformal or Bunch-Davies vacuum state respectively.  More importantly one can ask whether any such states exist which
are realistic for the universe that we live in.  This seems unlikely because today is certainly an early time compared to $t_r$ if our universe does have a big rip singularity in
its future; however the energy densities of quantized fields today are
much less than that of the dark energy.  Thus it would be necessary for the energy density of a quantum field
 to be small compared to the phantom energy density today and then to grow fast enough to become comparable to it well before the Planck scale is reached.
This type of behavior seems highly artificial, particularly since it is not what happens for a massless scalar field in the conformal or Bunch-Davies vacuum states.  Thus we find no evidence which would lead us to believe that backreaction effects due to quantum fields would remove a big rip singularity in our universe if, indeed, such a singularity lies in our future.

\acknowledgments

This work was supported in part by the National Science Foundation under Grant No.~PHY-0556292 and No.~PHY-0856050.
Numerical computations were performed on the Wake Forest University DEAC Cluster with
support from an IBM SUR grant and the Wake Forest University IS Department. Computational
results were supported by storage hardware awarded to Wake Forest University through
an IBM SUR grant.

\end{document}